\documentclass[showpacs]{revtex4}

\usepackage{amsmath}    
\usepackage{amssymb}
\usepackage{graphicx}   
\usepackage{verbatim}   
\usepackage{color}      
\usepackage{hyperref}   
\begin{document}

\title{Lovelock black holes in a string cloud background}
\author{Tae-Hun Lee$^{a\;}$}
\email{taehunee@gmail.com}
\author{Dharmanand Baboolal$^{a\;}$}
\email{Baboolald@ukzn.ac.za}
\author{Sushant G. Ghosh$^{a,\;b\;}$}
\email{sgghosh@gmail.com}
\affiliation{$^{a}$School of Mathematical Sciences, University of KwaZulu-Natal, Westville, Durban 4000, South Africa}
\affiliation{$^{b}$Centre for Theoretical Physics, Jamia Millia Islamia, New Delhi-110025, India}

\date{\today}

\begin{abstract}
We present an exact static, spherically symmetric black hole
solution to  the third order Lovelock gravity with a string cloud
background in seven dimensions for the special case when the second
and third order Lovelock coefficients are related via
$\tilde{\alpha}^2_2=3\tilde{\alpha}_3\;(\equiv\alpha^2)$. Further, we
examine thermodynamic properties of this black hole to obtain exact
expressions for mass, temperature, entropy and also perform the
thermodynamic stability analysis. We see that a string cloud
background makes a profound influence on horizon structure,
thermodynamic properties and the stability of black holes.
Interestingly the entropy of the black hole is unaffected due to a
string cloud background. However, the critical solution for
thermodynamic stability is being affected by a string cloud
background.
\end{abstract}



\maketitle

\section{Introduction}
Black holes through  quantum
outcome indicate that they radiate due to the Hawking effect
\cite{Hawking:1974rv}.
In the absence of established theories of quantum gravity,  black holes
have become a main playground to divulge quantum gravity effects through
their thermodynamics.  Black holes have been used as theorists'
laboratories in many other relevant fields. Thermodynamic
properties of black holes have been studied for many years, but
established statistical explanations of black hole thermodynamics
are still lacking. It shows that black holes also have the standard
thermodynamic quantities, such as temperature, entropy and heat capacity and so on, and
even possess abundant phase structures like the Hawking-Page phase
transition \cite{HPT} and similar critical phenomena in ordinary
thermodynamic systems.

Recent years witness the renewed interest towards the study of black hole
solutions specially in modified theories of gravity
\cite{Lovelock:1971yv,Buchdahl:1983zz}, as besides theoretical
results, cosmological evidence, e.g. dark matter and dark energy,
suggests possibility of changing the Einstein gravity. On the other
hand, the Einstein gravity cannot be quantized (non-renormalizable),
so it is believed that it is a low energy effective theory and could
be modified with higher derivative terms at high energy
\cite{Gross:1986iv}. Modifications to the Einstein gravity theory,
for instance the Lovelock theory \cite{Lovelock:1971yv}, the $f(R)$
gravity theory \cite{Buchdahl:1983zz}, etc. have been studied
extensively. Those models in higher spacetime dimensions have very
different features. For example, nature of stability in higher
dimensions is quite different. Extending spacetime dimensionality
in gravity theories has been one possible way to combine other
interactions with gravity or often seems to be even required in many
theories, e.g. Kaluza-Klein theory, a string theory, Brane world
scenarios, etc. \cite{Emparan:2008eg}.

In these context, apart from the standard Einstein-Hilbert action,
there also exist interesting theories of gravity in dimensions
greater than four involving higher powers of the curvatures such
that the field equations for the metric are at most in second-order.
Among the higher curvature gravity theories, the most extensively studied
theory is the so-called Lovelock gravity \cite{Lovelock:1971yv},
which naturally emerges when we wish to generalize the Einstein
theory in higher dimensions by keeping all characteristics of usual
general relativity excepting the linear dependence of the Riemann
tensor. The Lovelock gravity is one of the most general second order
theories in higher dimensions, which is free from ghosts.
The Lovelock theory may play an important role in a string theory
where the low energy effective field theory of gravity contains
higher curvature terms.

In this sense the Lovelock gravity \cite{Lovelock:1971yv} is a natural extension
to the Einstein gravity. It is constructed by sum of all the
Euler densities of a $2n$-dimensional manifold. The Lagrangian is given by
\begin{equation}
\mathcal{L}= \sum\limits_{n=0}^{t}\alpha _{n}\ \mathcal{L}_{n},
\end{equation}
where
\begin{equation}
\mathcal{L}_{n}=\frac{1}{2^{n}}\delta _{\alpha _{1}\beta_{1}...
\alpha _{n}\beta _{n}}^{\mu _{1}\nu _{1}...\mu _{n}\nu_{n}}
\prod\limits_{r=1}^{n}R_{\quad \mu _{r}\nu _{r}}^{\alpha _{r}\beta _{r}}.
\end{equation}
A spacetime dimension $D$ can be written as $D=2t+2$ for even
dimensions and  $D=2t+1$ for odd dimensions. The $n$th order higher
derivative terms $\mathcal{L}_{n}$ become a surface term when $D\leq
2n$. Non-trivial extra terms contribute to equations of motion in
higher dimensions but not in dimensions less than $2n$. Moreover,
higher derivative terms can cancel a ghost term. For instance, the
reference \cite{Zwiebach:1985uq} shows that the second order
Lovelock (Gauss-Bonnet) terms cancel a ghost term. Boulware and Deser  \cite{bd}  first found a static, spherically symmetric black
hole solution with the Gauss-Bonnet corrections. Using the
Gauss-Bonnet gravity static, spherically symmetric solutions are
obtained later \cite{Wheeler:1985qd, Wheeler:1985nh} with thermodynamic properties
\cite{Myers:1988ze}. Static, spherically symmetric black hole
solutions in the Lovelock gravity with general energy momentum
tensors in any arbitrary dimension can be found in
\cite{Mazharimousavi:2008ti} and \cite{LLP} and its thermodynamics in \cite{Cai:2003kt}. Further extensive studies on the Gauss-Bonnet black holes with a focus on thermodynamic properties have been found in \cite{Cai:2001dz, Cai:2003gr}. The special third order Lovelock gravity also received a significant attention, e.g.
for a black hole solution and its thermodynamics in this theory with Born-Infield source \cite{Li:2011uk, Camanho:2011rj} and for a causality violation \cite{Camanho:2014apa}. Also, topological properties of the general Lovelock black holes in the context of thermodynamics have been investigated \cite{Cai:1998vy}.

In this paper, we begin with finding static, spherically symmetric
black hole solutions for a string cloud background for a specific case,
i.e., $\tilde{\alpha}^2_2=3\tilde{\alpha}_3$ and examine
thermodynamic properties  in the third order
Lovelock gravity.  The solution in there can be utilized to
calculate mass, temperature, entropy and heat capacity of black
holes and explicitly study effects of a string cloud background. It turns out that the horizon and thermodynamic properties of Lovelock black hole in
 conjunction with a string parameter could have some
interesting features. It may be pointed out that gravity coupled to
clouds of strings may be very useful and important as the Universe
can be considered as a collection of extended objects, like, one
dimensional strings. The study of black holes
in a cloud of strings was initiated by Letelier modifying the Schwarzschild solutions for a cloud of strings as a source \cite{Letelier:1979ej}, which was
recently extended to the Gauss-Bonnet gravity
\cite{Herscovich:2010vr} and also to the Lovelock gravity
\cite{sgg_sdm, Ghosh:2014pga}. We show that a string cloud background makes a profound influence on horizon structures and thermodynamic quantities but entropy is not
changed.

The paper is organized as follows: In Sec.~\ref{Lovelock action and
Equations of Motion}  we begin examining the third order Lovelock
action,  which is a modification of the Einstein-Hilbert action, and
also derive energy momentum tensors of a cloud of strings. The
thermodynamics of a static, spherically symmetric black hole solution in
this theory is explored in Sec.~\ref{Thermodynamics of black holes}.
Before that we find an exact static, spherically symmetric black hole
solution in Sec.~\ref{Spherically Symmetric  Solution in Lovelock
Gravity}. The paper ends in Sec.~\ref{Conclusion}, which gives
concluding remarks. We have used units that fix $G=c=1$ and the metric
signature, $(-,+,+,\cdots,+)$.

\section{Lovelock action and Equations of Motion}\label{Lovelock action and Equations of Motion}
The Lovelock theory is the most general theory of gravity that gives second order field equations in arbitrary dimensions. The recent interest in the Lovelock theory arose because its action appears as a low energy limit of a heterotic superstring theory.  The simplest third order Lovelock action reads \cite{Lovelock:1971yv}:
\begin{equation}
\mathcal{I}_{G}=\frac{1}{2}\int_{\mathcal{M}}dx^{D}\sqrt{-g}\left[  \mathcal{L}_{1} +\alpha_{2}\mathcal{L}_{GB}+\alpha_{3}\mathcal{L}_{\left(  3\right)
} \right] + \mathcal{I}_{S} \label{IG}
\end{equation}
where $\mathcal{I}_{S}$ is a matter action due to cloud of strings. The Einstein term $ \mathcal{L}_1$ is $R$, the second order Lovelock (Gauss-Bonnet) term $\mathcal{L}_{GB}$  is
\begin{equation}
\mathcal{L}_{GB}=R_{\mu\nu\gamma\delta}R^{\mu
\nu\gamma\delta}-4R_{\mu\nu}R^{\mu\nu}+R^{2}
\end{equation}
and the third order Lovelock Lagrangian is
\begin{align}
\mathcal{L}_{\left(  3\right)  }  &  =2R^{\mu\nu\sigma\kappa}R_{\sigma
\kappa\rho\tau}R_{\quad\mu\nu}^{\rho\tau}+8R_{\quad\sigma\rho}^{\mu\nu
}R_{\quad\nu\tau}^{\sigma\kappa}R_{\quad\mu\kappa}^{\rho\tau}\nonumber\\
&  +24R^{\mu\nu\sigma\kappa}R_{\sigma\kappa\nu\rho}R_{\ \mu}^{\rho}%
+3RR^{\mu\nu\sigma\kappa}R_{\sigma\kappa\mu\nu}\nonumber\\
&  +24R^{\mu\nu\sigma\kappa}R_{\sigma\mu}R_{\kappa\nu}+16R^{\mu\nu}%
R_{\nu\sigma}R_{\ \mu}^{\sigma}\\
&  -12RR^{\mu\nu}R_{\mu\nu}+R^{3}. \nonumber
\end{align}
Here $R,$ $R_{\mu\nu\gamma\delta\text{ }}$ and $R_{\mu\nu}$ are the
Ricci scalar, the Riemann and the Ricci tensors, respectively. The
coupling constants $\alpha_2$ and $\alpha_3$ in Eq.~(\ref{IG}) have
dimensions, $[\text{length}]^{4-D}$ and $[\text{length}]^{6-D}$,
respectively and will help us to explicitly bring out changes in the general relativity equations. In the limits
$(\alpha_2,\alpha_3) \rightarrow 0$, one recovers the
Einstein-Hilbert action. The variation of the action with respect to the
metric $g_{\mu\nu}$ yields modified field equations for the third
order Lovelock gravity,
\begin{equation}
G_{\mu\nu}^{E}+\alpha_{2}G_{\mu\nu}^{GB}+\alpha_{3}G_{\mu\nu}^{\left(
3\right)  }=T_{\mu\nu},\label{ee}
\end{equation}
where
$G_{\mu\nu}^{E}$ is the Einstein tensor, while $G_{\mu\nu}^{GB}$ and
$G_{\mu\nu}^{\left(  3\right)  }$ are given explicitly in \cite{zz}, respectively :
\begin{eqnarray}
 G_{\mu\nu}^{GB} & = & 2\;\Big( -R_{\mu\sigma\kappa\tau}R_{\quad\nu}^{\kappa
\tau\sigma}-2R_{\mu\rho\nu\sigma}R^{\rho\sigma}-2R_{\mu\sigma}R_{\ \nu
}^{\sigma} \nonumber \\ & &  +RR_{\mu\nu}\Big) -\frac{1}{2}g_{\mu\nu}\mathcal{L} _{GB},
\end{eqnarray}
and
\begin{gather}
G_{\mu\nu}^{\left(  3\right)  }=-3\left(  4R_{\qquad}^{\tau\rho\sigma\kappa
}R_{\sigma\kappa\lambda\rho}R_{~\nu\tau\mu}^{\lambda}-8R_{\quad\lambda\sigma
}^{\tau\rho}R_{\quad\tau\mu}^{\sigma\kappa}R_{~\nu\rho\kappa}^{\lambda}\right. \nonumber
\\
+2R_{\nu}^{\ \tau\sigma\kappa}R_{\sigma\kappa\lambda\rho}R_{\quad\tau\mu
}^{\lambda\rho}-R_{\qquad}^{\tau\rho\sigma\kappa}R_{\sigma\kappa\tau\rho
}R_{\nu\mu}+8R_{\ \nu\sigma\rho}^{\tau}R_{\quad\tau\mu}^{\sigma\kappa
}R_{\ \kappa}^{\rho}\nonumber\\
+8R_{\ \nu\tau\kappa}^{\sigma}R_{\quad\sigma\mu}^{\tau\rho}R_{\ \rho}^{\kappa
}+4R_{\nu}^{\ \tau\sigma\kappa}R_{\sigma\kappa\mu\rho}R_{\ \tau}^{\rho
}-4R_{\nu}^{\ \tau\sigma\kappa}R_{\sigma\kappa\tau\rho}R_{\ \mu}^{\rho
}\nonumber\\
+4R_{\qquad}^{\tau\rho\sigma\kappa}R_{\sigma\kappa\tau\mu}R_{\nu\rho}%
+2RR_{\nu}^{\ \kappa\tau\rho}R_{\tau\rho\kappa\mu}+8R_{\ \nu\mu\rho}^{\tau
}R_{\ \sigma}^{\rho}R_{\ \tau}^{\rho}\nonumber\\
-8R_{\ \nu\tau\rho}^{\sigma}R_{\ \sigma}^{\tau}R_{\ \mu}^{\rho}-8R_{\quad
\sigma\mu}^{\tau\rho}R_{\ \tau}^{\sigma}R_{\nu\rho}-4RR_{\ \nu\mu\rho}^{\tau
}R_{\ \tau}^{\rho}\nonumber\\
+4R_{\quad}^{\tau\rho}R_{\rho\tau}R_{\nu\mu}-8R_{\ \nu}^{\tau}R_{\tau\rho
}R_{\ \mu}^{\rho}+4RR_{\nu\rho}R_{\ \mu}^{\rho}
\left.  -R^{2}R_{\nu\mu}\right) \nonumber \\ -\frac{1}{2}g_{\mu\nu}\mathcal{L}_{(3)}.\nonumber
\end{gather}
and $T_{\mu\nu}$ is the energy momentum tensor of a matter field which we consider here as clouds of strings.  Note that for the third order Lovelock gravity, the
non-trivial third terms require a spacetime dimension $D$ to satisfy $ D \geq 7 $.
\subsection{Energy Momentum Tensor}
Next we turn attention to calculate the energy-momentum tensor of a cloud of strings (see \cite{Letelier:1979ej}, for further details). The Nambu-Goto action of a string evolving in spacetime is given by
\begin{equation}\label{saction}
\mathcal{I}_S = \int_{\Sigma} \mathcal{L} \;  d\lambda^{0} d\lambda^{1}, \hspace{0.2in} \mathcal{L} = m (\gamma)^{-1/2},
\end{equation}
with the Lagrangian for a cloud of strings \cite{Letelier:1979ej}:
\[     \mathcal{L} = m \left[-\frac{1}{2} \Sigma^{\mu \nu} \Sigma_{\mu \nu}\right]^{1/2}.\]
The string worldsheet is associated with a bivector of the form
\begin{equation}
\label{eq:bivector}
     \Sigma^{\mu \nu} = \epsilon^{a b} \frac{\partial x^{\mu}}{\partial \lambda^{a}} \frac{\partial x^{\nu}}{\partial \lambda^{b}},
\end{equation}
where $\epsilon^{a b}$ is the two-dimensional Levi-Civita tensor and  $\epsilon^{0 1} = - \epsilon^{1 0} = 1$.
Here $m>0$ is a constant for each string and $\gamma$ is the determinant of an induced metric on the string world sheet $\Sigma$   given by
\begin{equation}
 \gamma_{a b} = g_{\mu \nu} \frac{\partial x^{\mu}}{\partial \lambda^{a}} \frac{\partial x^{\nu}}{\partial \lambda^{b}}.
\end{equation}
$(\lambda^{0}, \lambda^{1})$ with $\lambda^{0}$ and $\lambda^{1}$ which are a timelike and a spacelike parameter, respectively, is a parametrization of the world sheet $\Sigma$, \cite{synge}.
Further, since $T^{\mu \nu} = -2 \partial \mathcal{L}/\partial g^{\mu \nu}$, then the energy-momentum tensor for one string reads
\begin{equation}
T^{\mu \nu} = m \Sigma^{\mu \rho} \Sigma_{\rho}^{\phantom{\rho} \nu}/(-\gamma)^{1/2}.
\end{equation}
Hence, a cloud of strings has the energy-momentum tensor
\begin{equation}
T^{\mu \nu} = \rho {\Sigma^{\mu \sigma} \Sigma_{\sigma}^{\phantom{\sigma} \nu}}/{(-\gamma)^{1/2}  },\label{EM}
\end{equation}
where $\rho$ is a proper density of a cloud of strings. The quantity $\rho \; (-\gamma)^{-1/2} $ is a gauge-invariant density.

\section{Effect of String Cloud}
As will be seen in next discussions the presence of a string cloud plays a main role in the horizon structure of the theory together with other parameters. For instance, a string cloud parameter $a$ can change the number of horizons and singularities and make singularity covered by a horizon with fixed other parameters. Given parameters $(\alpha,a)$ a positive mass condition imposes either mass bound or range of the horizon radius. Rich analysis along this line for absence of energy momentum tensor has been made in \cite{Camanho:2011rj}. Vanishing string cloud is a transition point for singularities to be created $\alpha>0$ and one for the number of horizons for $\alpha<0$. In general the energy momentum tensor $T^\mu_\nu$ is expressed for static spherically symmetric spacetime,
\begin{equation}
T^\mu_{\nu}=\frac{a}{r^{n(1-k)}}\text{diag}(1,1,k,\cdots, k),\label{energy_momentum}
\end{equation}
where $k$ is a constant, \cite{Salgado:2003ub}.
The dominant energy condition allows only $a\leq0$ and $-1\leq k\leq0$ and the causality condition further constrains $k$ to $-1\leq k\leq-\frac{1}{n}$ \cite{Salgado:2003ub}. Thermodynamic quantities are determined by mass expression, which is given for the Lovelock theory in terms of a horizon radius $r_h$ by
\begin{equation}
M=\xi\sum\limits^m_{s=0}\tilde{\alpha}_s\kappa^sr^{-2s+n+1}_h+\xi \frac{2}{n}\int^{r_h}  r^nT^t_t dr,\label{M}
\end{equation}
where $\xi\equiv\frac{n}{16\pi}2\pi^{(n+1)/2}/\Gamma[(n + 1)/2]$ and the curvature constant $\kappa=-1,0,1$, \cite{THLee:2015}. The mass for a string cloud is
\begin{equation}
M_{sc}=\xi\sum\limits^m_{s=0}\tilde{\alpha}_s\kappa^sr^{-2s+n+1}_h+\xi\frac{2a}{n}r_h.\label{Ms}
\end{equation}
As noticed the a string cloud contributes to the mass with the highest possible power of $r_h$, i.e. the energy condition upper bound $k=0$ but violates causality. For even dimensions the presence of a string cloud effectively changes the highest coupling $\alpha_m$, $m=2n$ while for odd dimensions it is a unique source term thermodynamically, \cite{THLee:2015}.

In this paper we take a string cloud only as a classical background. The reference \cite{Camanho:2014apa} claims that causality violation in the third order Lovelock theory can be fixed by adding an infinite tower of massive higher spin particles. Unless we look for dynamics instead of static background of strings, it is not certain how a string cloud contributes to the causality problem. However, it is worthwhile to study whether causality violation in classical sense can be weaken or removed when the external massive particles are combined with a string cloud into a background.

\section{Spherically Symmetric  Solution in Lovelock Gravity}\label{Spherically Symmetric  Solution in Lovelock Gravity}
Here we wish to obtain static, spherically symmetric black hole solutions to Eq.~(\ref{ee}) for the energy momentum tensors, Eq.~(\ref{EM}), and investigate their horizons and thermodynamic properties. Hence, we assume the metric of the form:
\begin{equation}\label{metric}
ds^2=-f(r) dt^2+ \frac{1}{f(r)} dr^2 + r^2 \tilde{\gamma}_{ij}\; dx^i\; dx^j,
\end{equation}
where $ \tilde{\gamma}_{ij} $ is a metric of a $(D-2)$-dimensional constant curvature space $\kappa = 1,\; 0 \;$ or -1, representing spherical, flat and hyperbolic spaces, respectively. But, in this paper we shall confine ourselves to $\kappa = 1$.  To find the metric function $f(r)$, we should solve Eq.~(\ref{ee}). Using this metric ansatz, an $r-r$ component of the field  equations of motion reduces to:
\begin{eqnarray}\label{master}
& &
\left[r^{5}-2\tilde{\alpha}_{2}r^{3}\left(  f\left(  r\right)  -1\right)
+3\tilde{\alpha}_{3}r\left(  f\left(  r\right)  -1\right)  ^{2}\right]
f^{\prime}\left(  r\right)  +\nonumber\\
& & \left(  n-1\right)  r^{4}\left(  f\left(  r\right)  -1\right)  -\left(
n-3\right)  \tilde{\alpha}_{2}r^{2}\left(  f\left(  r\right)  -1\right)
^{2}+\nonumber\\
& & \left(  n-5\right)  \tilde{\alpha}_{3}\left(  f\left(  r\right)  -1\right)^{3} = \frac{2r^6 }{n}T^r_r, \label{EGBL rr}
\end{eqnarray}
where a prime denotes a derivative with respect to $r,$ $n\equiv D-2,$
$\tilde{\alpha}_{2}\equiv(n-1)(n-2)\alpha_{2}$ and
$\tilde{\alpha}_{3}\equiv(n-1)(n-2)(n-3)(n-4)\alpha_{3}$.
 In general, the Eq.~(\ref{master}) has one real and two complex solutions. But it can also have three real solutions as well under appropriate conditions. We are seeking static, spherically symmetric real solutions, which restrict a density $\rho$ and a bivector $\Sigma_{\mu \nu}$ as a function of $r$ only.  Further, the only possible non-zero component of a bivector $\Sigma$ is $\Sigma^{tr} =  - \Sigma^{rt}$.  Thus $T^t_t=T^r_r=-\rho \Sigma^{tr}$ and we obtain  $\partial_{r} (\sqrt{r^n T^t_t}) = 0$,  which implies:
 \begin{equation}
 T^t_t = T^r_r = \frac{a}{r^n},
 \end{equation}
 for some real constant $a$. Clearly the third order Lovelock gravity is non-trivial only for spacetime dimensions $D \geq 7$.  Henceforth, to extract information from our analysis, we shall confine ourselves to $D=7$ in which the  Eq.~(\ref{EGBL rr}) can be easily integrated and the solution reads
\begin{equation}
f(r)=1+\frac{\tilde{\alpha} _{2} }{3 \tilde{\alpha} _{3}}r^2
+\frac{10^{1/3} \left(\tilde{\alpha} _{2}^2-3 \tilde{\alpha} _{3}\right)}{3 \tilde{\alpha} _{3}A^{1/3}} r^6 +B^{1/3},
\end{equation}
where
\begin{equation}
A\equiv5\tilde{\alpha} _{2}(2\tilde{\alpha} _{2}^2 -9 \tilde{\alpha} _{3})r^6-54 \tilde{\alpha} _{3}^2 \left(a r+C_1\right)+3\tilde{\alpha} _{3}(3\Delta)^{1/2},
\end{equation}
and
\begin{equation}
B\equiv\frac{1}{3^3\times 10^{1/3} \tilde{\alpha} _{3}}[5\tilde{\alpha} _{2}(2 \tilde{\alpha} _{2}^2 -9 \tilde{\alpha} _{3})r^6 +54 \tilde{\alpha} _{3}^2(a r-C_1)+3\tilde{\alpha} _{3}(3\Delta)^{1/2}],
\end{equation}
with
\begin{equation}
\Delta\equiv-25 \left(\tilde{\alpha} _{2}^2-4 \tilde{\alpha} _{3}\right) r^{12}+20 \tilde{\alpha} _{2} \left(2 \tilde{\alpha} _{2}^2-9 \tilde{\alpha} _{3}\right) r^6 \left(a r-C_1\right)+108 \tilde{\alpha} _{3}^2 \left(C_1-a r\right)^2.
\end{equation}
Here we shall impose the condition $\tilde{\alpha} _{2}^2=3\tilde{\alpha} _{3}\;(\equiv\alpha^2)$, which simplifies $f(r)$ significantly with all coefficients intact and hence it is worthwhile to consider this case. This condition reduces $f(r)$ to the following form
\begin{equation}
f(r)= 1+ \frac{r^2}{\alpha} +
\frac{1}{2^{1/3}\alpha^2}\left\{-\alpha^3 g(r)+\alpha^2\sqrt{[-\alpha g(r)]^2}\right\}^{1/3},\label{f*}
\end{equation}
where $g(r)$ is given with the notation $\omega\equiv \frac{2}{5}C_1$,
\begin{equation}
g(r)=r^6-\frac{6}{5}a\alpha r+3\omega \alpha.
\end{equation}

\begin{figure}[h!]
  \begin{center}
\begin{tabular}{ccc}
    \includegraphics[scale=1.0]{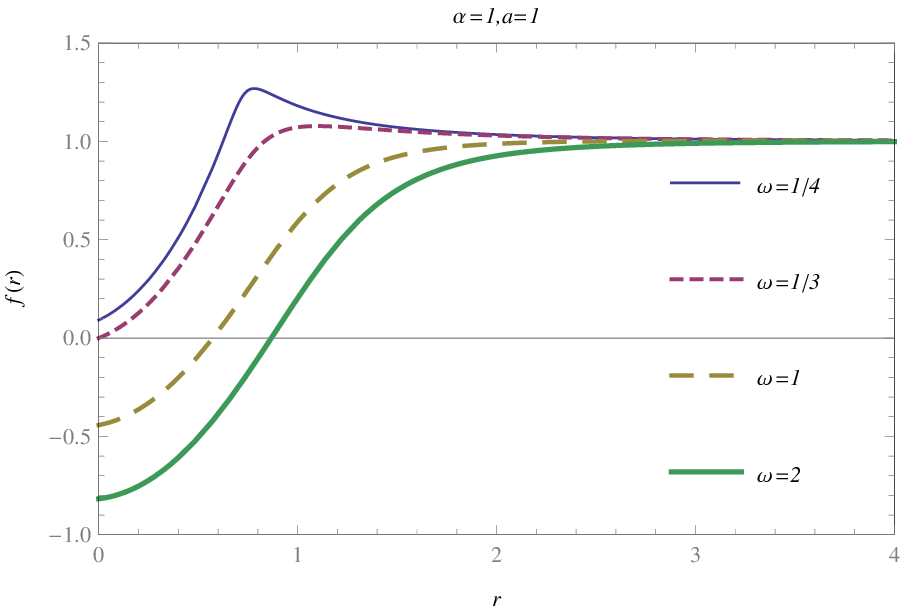} &&
    \includegraphics[scale=1.0]{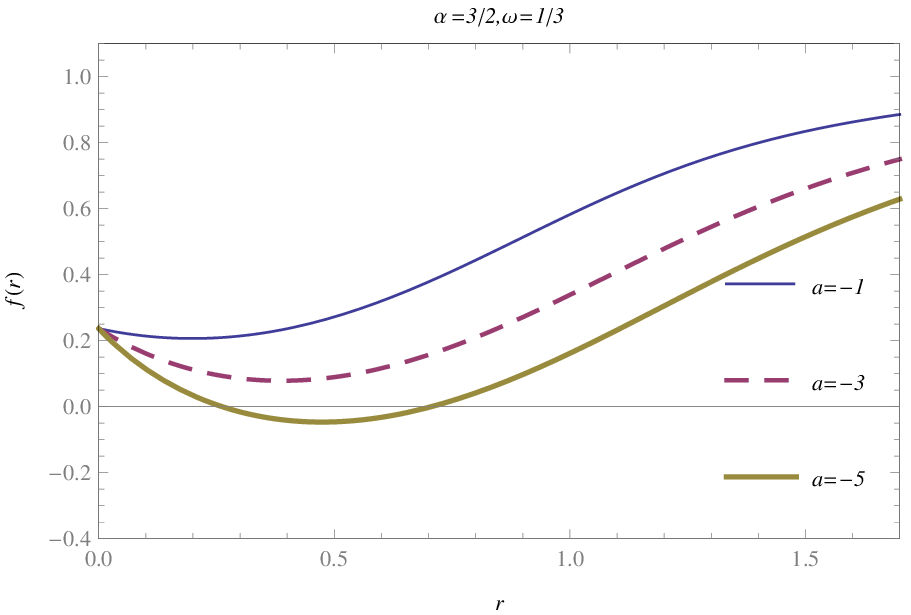}\\
    (a)&&(b)\\
    \includegraphics[scale=1.0]{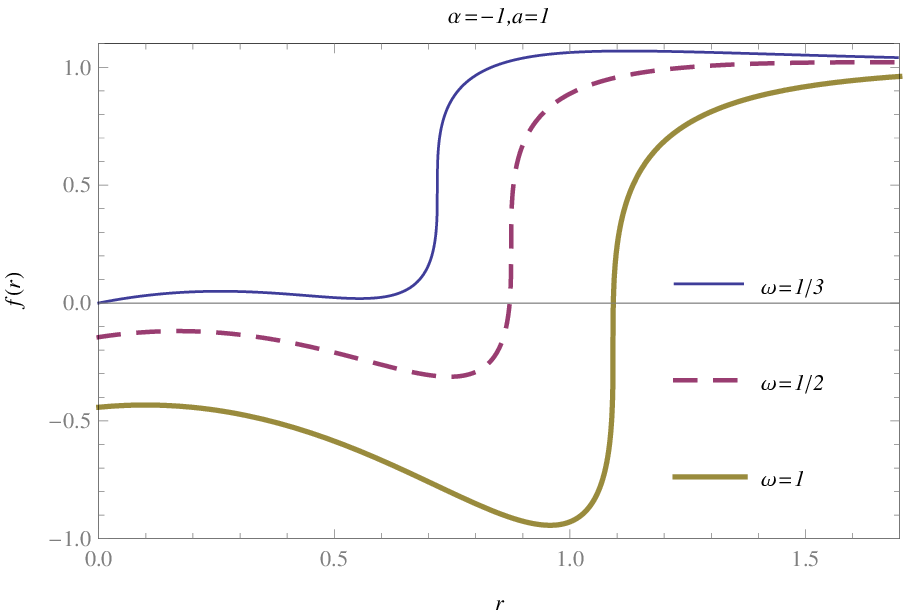} &&
    \includegraphics[scale=1.0]{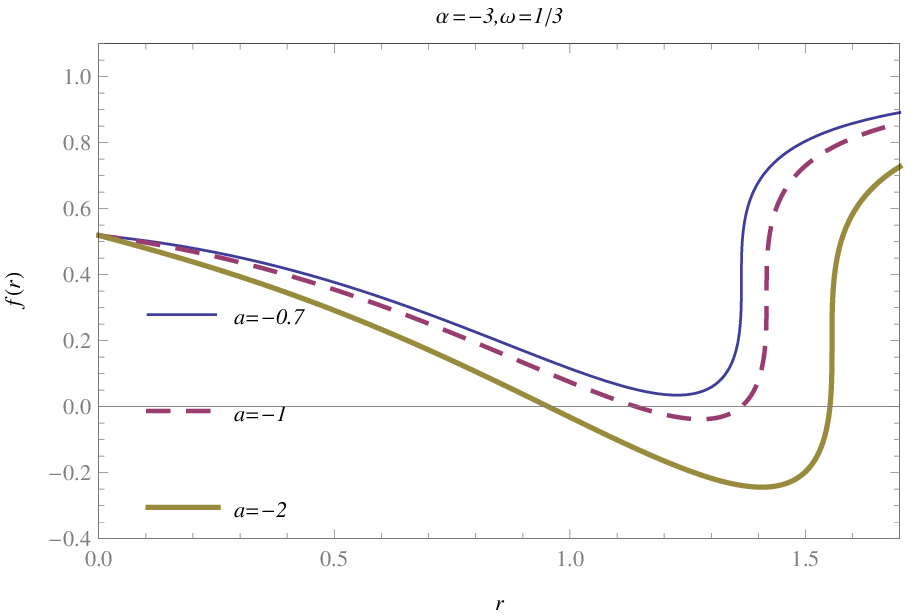}\\
    (c)&&(d)
\end{tabular}
    \end{center}
    \caption{Plots show a metric function $f(r)$ as a function of $r$ for different values of parameters.}\label{fig:f}
\end{figure}

Note that the square root here should be defined including a sign of
$-\alpha g(r)$, i.e., $\sqrt{[-\alpha g(r)]^2}=-\alpha g(r)$. One can
confirm it by noticing that the solution has been obtained from
solving a cubic equation, \cite{Wheeler:1985qd}. Eq.~(\ref{EGBL rr}) can be rewritten as
\begin{equation}
-r^{-n+6}\frac{\partial}{\partial
r}[r^{n-1}(r^2F)+\tilde{\alpha}_2r^{n-3}(r^2F)^2+\tilde{\alpha}_3r^{n-5}(r^2F)^3]=r^{-n+6}\frac{\partial}{\partial
r}\int dr\frac{2r^n}{n}T^r_r(r),
\end{equation}
where $F\equiv[1-f(r)]/r^2$. For $n=5$ and
$\tilde{\alpha}^2_2=3\tilde{\alpha}_3=\alpha^2$ this reduces to,
with an integration constant $\omega$,
\begin{equation}
(\alpha
F+1)^3=1-\frac{6a\alpha}{5r^5}+\frac{3\omega\alpha}{r^{6}}=\frac{g}{r^{6}}.
\end{equation}
Thus $f(r)$ reads:
\begin{equation}
f(r)=1+\frac{r^2}{\alpha}\left[1-\left(1-\frac{6}{5r^5}a\alpha+\frac{3}{r^6}\omega\alpha\right)^{1/3}\right].\label{f-}
\end{equation}
We consider only positive mass, i.e., $\omega\geq0$.  $f(r)$
asymptotically behaves as $\lim\limits_{r\to\infty}f(r)=1$ as shown
in FIG.~\ref{fig:f} with various parameter values.

In fact, for $D > 4$ the Einstein gravity can be thought of as a particular case of the Lovelock gravity since the Einstein-Hilbert term is one of several terms that constitute the Lovelock action. Hence, for  $D > 4$ and $\alpha = 0 $, the higher dimensional Schwarzschild solution in a string cloud model reads \cite{sgg_sdm}:
\begin{equation}
f\left(r\right)  =1-\frac{\omega}{r^{D-3}}+ \frac{2a}{(D-2)r^{D-4}},\label{HDSchwarzchild}
\end{equation}
which goes to Eq.~(\ref{fasym}) for $D=7$.
Eq.~(\ref{f-}), in the limit $\alpha\to 0$, again leads to
\begin{equation}
 f(r)=1-\frac{\omega}{r^4} +\frac{2}{5r^3}a+O(\alpha).\label{fasym}
\end{equation}
which can be also obtained from Eq.~(\ref{HDSchwarzchild}) when
$D=7$. With $a=0$, $f(r)$ in Eq.~(\ref{f-}) can be
identified with one in the reference \cite{Li:2011uk} with
$\beta=0$, i.e. the vanishing Born-Infield field. The solution can
be exactly verified through the reference
\cite{Mazharimousavi:2008ti}. Since $\omega$ represents mass it
should be positive, $\omega\geq 0$.

Due to a fractional power on $g(r)$, $f(r)=1+\frac{r^2}{\alpha}-\frac{g^{1/3}}{\alpha}$ leads to a curvature singularity at $r=r_*$, where $g(r_*)=0$ as in the Gauss-Bonnet case \cite{Myers:1988ze}. Using the expression of the Ricci scalar $R=-[(n^2+2)F+(n+4)rF^\prime/2+r^2F^{\prime\prime}]$ from \cite{Wheeler:1985qd}, one sees $\lim\limits_{r\to r_*}R=\infty$. A horizon radius $r_h$ is defined by $f(r_h)=0$.

To see how $r_h$ and $r_*$ can be related, it is convenient to define $D(r)\equiv(r^2+\alpha)^3-g(r)=3\alpha(r^4_h+\alpha r^2_h+\frac{2}{5}ar_h-\omega+\frac{1}{3}\alpha^2)$, where $g(r)=r^6-\frac{6}{5}a\alpha r+3\omega\alpha$. Note that $0=D(r_h)=(r_h^2+\alpha)^3-g(r_h)$ and $D(r_*)=(r_*^2+\alpha)^3$.
Let us first consider $\alpha>0$ and $a>0$ case. It is worthwhile to notice that $g(r_h)>0$ and $\frac{\partial D}{\partial r}=6\alpha(2r^3+\alpha r+\frac{1}{5}a)>0$. The latter tells us that only one horizon can exist. A necessary and sufficient condition for existence of one and only one horizon can be found to be a negative $y$-intercept, i.e., $\frac{1}{3}\alpha^2-\omega\leq0$. $g$ has a minimum at $r=(a\alpha/5)^{1/5}$. In case that $g$ has a negative minimum, hence giving two singularities, $(r_{*s}, r_{*b})$, a horizon can stay in either $0<r_h< r_{*s}$ or $r_{*b}<r_h$ due to $g(r_h)>0$, but the case $r_{*b}<r_h$ leads to $D(r_{*b})<0$, which is not true. This means that a horizon can exist only in $0<r_h< r_{*s}$, not being covered by a horizon.

For $\alpha>0$ and $a<0$ case $D(r)$ has one minimum, which can lead to two horizons if $\frac{1}{3}\alpha^2-\omega\geq0$. For $a\leq0$ $g$ is a monotonously increasing function with a positive a $y$-intercept and hence there is no singularity.

When $a=0$ there exists one horizon if $\frac{1}{3}\alpha^2-\omega\leq0$ and is no singularity.

In short for $\alpha>0$, $\frac{1}{3}\alpha^2-\omega<0$ is a necessary and sufficient condition for existence of one and only one horizon for all $a$. Non-negative $a$ allows only one horizon while negative $a$ two horizons at most, whose necessary condition is $\frac{1}{3}\alpha^2-\omega\geq0$. Positive $a$ gives a naked singularity while $a<0$ does not give a singularity. Without a string cloud $(a=0)$ there is no singularity.

Next, consider $\alpha<0$ case. There is one and only one singularity in this case. It is useful to notice that a horizon cannot exists between $r_0(\equiv |\alpha|^{1/2})$ and $r_*$. This can be checked by $D(r)>0$ for $r_0<r<r_*$ and $D(r)<0$ for $r_*<r<r_0$. We are particularly interested in whether singularities are covered by horizons. Because horizons cannot stay between $r_0$ and $r_*$ as mentioned above, the inequality between $r_0$ and $r_h$ is equivalent to one between $r_*$ and $r_h$, i.e. the criteria for whether singularities are covered by horizons or not. The solutions for horizons are intercepts between the curve $l_1=r^4+\alpha r^2$ and the straight line $l_2=-\frac{2}{5}ar+\omega-\frac{1}{3}\alpha^2$.

For $a>0$, $\frac{\partial D}{\partial r}$ can have at most two zeros, leading up to maximum three horizons. If a singularity is covered by a horizon, the horizon exists in $r>r_0$ and is the only one.

For $a<0$ $\frac{\partial D}{\partial r}$ has one and only one zero, leading to maximum two zeros in $D$. Let us consider possibility to have two intercepts for $r>r_0$. The necessary conditions are $a<0$ and $\omega-r_0^4/3\leq 2ar_0/5$. The slopes of $l_1$ and $l_2$ are equal to $-2a/5$ when they have one intercept. Thus $-2a/5>\partial l_1/\partial r\vert_{r=r_0}$, i.e., $-a\geq5r^3_0$. However, the condition $\omega\leq 2ar_0/5+r_0^4/3$ with the first one $-a\geq 5r^3_0$ conflicts with the positive mass condition $\omega\geq0$, i.e., $\omega\leq 2a r_0/5+r_0^4/3\leq -2r^4_0+r^4_0/3=-5r^4_0/3$. Therefore, there can exist only one horizon in the region $r>r_0$ and the condition for the existence of one horizon is $\omega> r_0^4/3+2ar_0/5$. It is important to notice that this condition is just $g(r_0)<0$. Once one horizon exists in $r>r_0$, the other exists in $r_{hs}<r_0$ and a singularity exists in $r_0<r_*<r_{hb}$, i.e. a singularity stays closer to the greater horizon $r_{hb}$.

For $a=0$  there is a similar property to case $a>0$, i.e. there are at most two horizons. Only when a singularity is covered by a horizon, it is the only horizon.

The sign of $\l_2(r_0)$, equal to $-3r_0^2g(r_0)$, tells us both whether $r_h$ is ahead or behind from $r_0$ and whether $r_*$ is ahead or behind from $r_0$. Therefore, along with the fact that there is no horizon between $r_0$ and $r_*$, only configuration, $r_0\geq r_*\geq r_h$ or $r_h\geq r_*\geq r_0$ is allowed, in other words, $r_*$ stays closer to $r_h$ than $r_0$.

In short for $\alpha<0$, there is only one singularity. For $a\geq0$ when a singularity is non-naked there should be only one horizon. For $a<0$, when a singularity is non-naked, the singularity must be between two horizons.

\begin{figure}[h!]
  \begin{center}
\begin{tabular}{ccc}
    \includegraphics[scale=1.0]{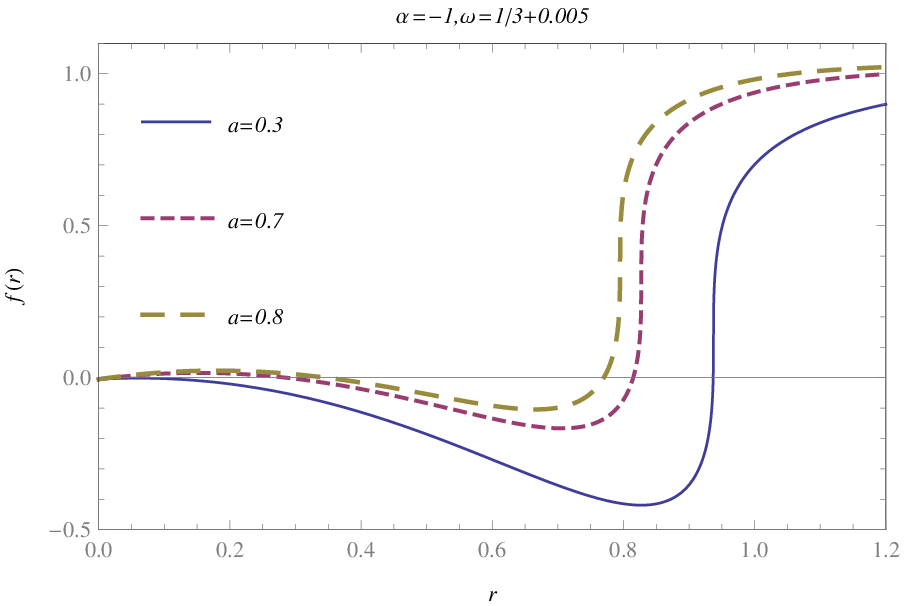} &&
    \includegraphics[scale=1.0]{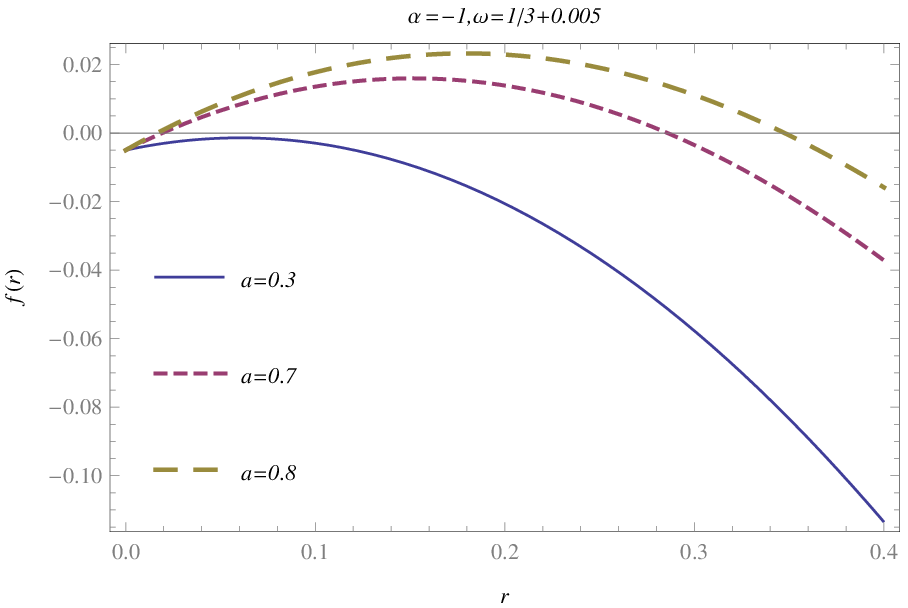}\\
    (a)&&(b)
\end{tabular}
    \end{center}
    \caption{Plots show that for $a>0$, $\alpha<0$ and $\omega-\alpha^2/3>0$ there can exist maximum three horizons in the region $r<r_*$.}\label{fig:h3}
\end{figure}

In the case that a singularity is covered by a horizon, the condition $\omega-2ar_0/5-r_0^4/3\geq0$ must be satisfied and it gives the lower bound of mass, unless $\frac{2}{5}ar_0+\frac{1}{3}r^4_0<0$,

\begin{equation}
\omega_{m} \equiv\frac{2}{5}ar_0+\frac{1}{3}r^4_0.\label{omegam}
\end{equation}
If $\omega_{m}<0$ the lower bound for $\omega$ must be taken to be zero.
\begin{figure}[h!]
  \begin{center}
    \includegraphics[scale=1.2]{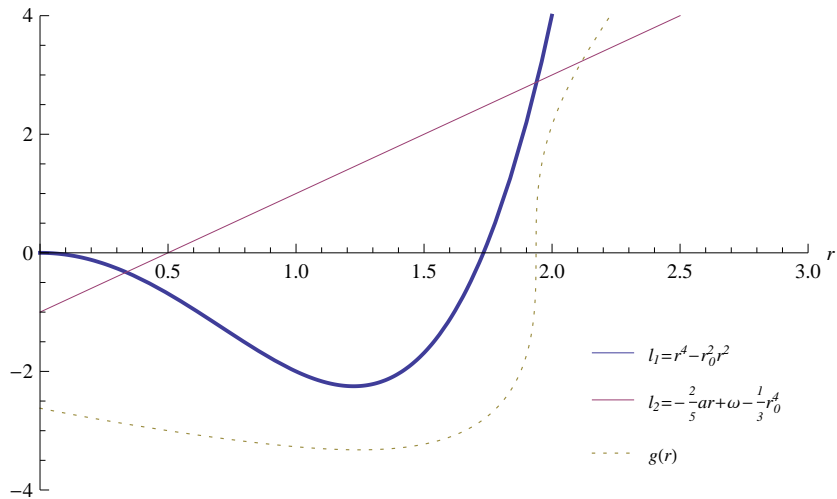}
    \end{center}
    \caption{The plot illustrates that horizons can be obtained from intercepts between $l_1=r^4-r^2_0r^2$ and $l_2=-\frac{2}{5}ar+\omega-\frac{1}{3}r^4_0$}\label{fig:h}
\end{figure}
Fig.~\ref{fig:f} shows behaviours of the metric function $f(r)$ with different values of parameters.
\section{Thermodynamics of black holes}\label{Thermodynamics of black holes}

\begin{figure}[h!]
  \begin{center}
\begin{tabular}{ccc}
    \includegraphics[scale=1.0]{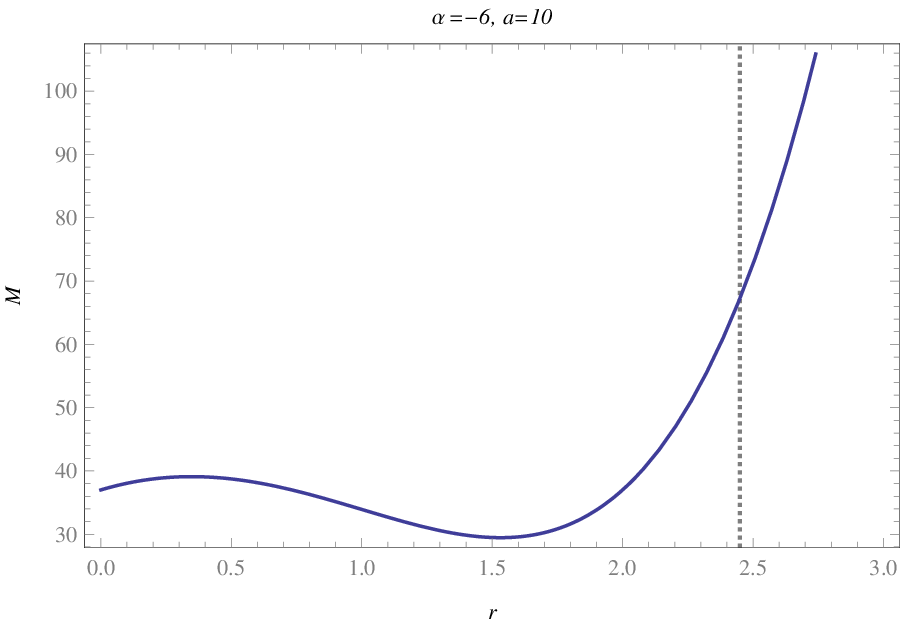} &&
    \includegraphics[scale=1.0]{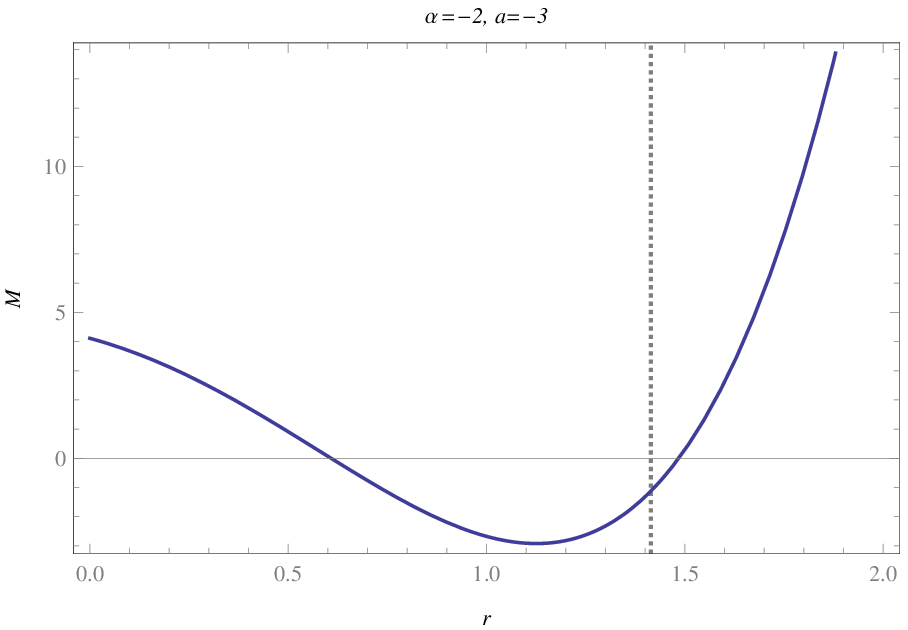}\\
    (a)&&(b)\\
    \includegraphics[scale=1.0]{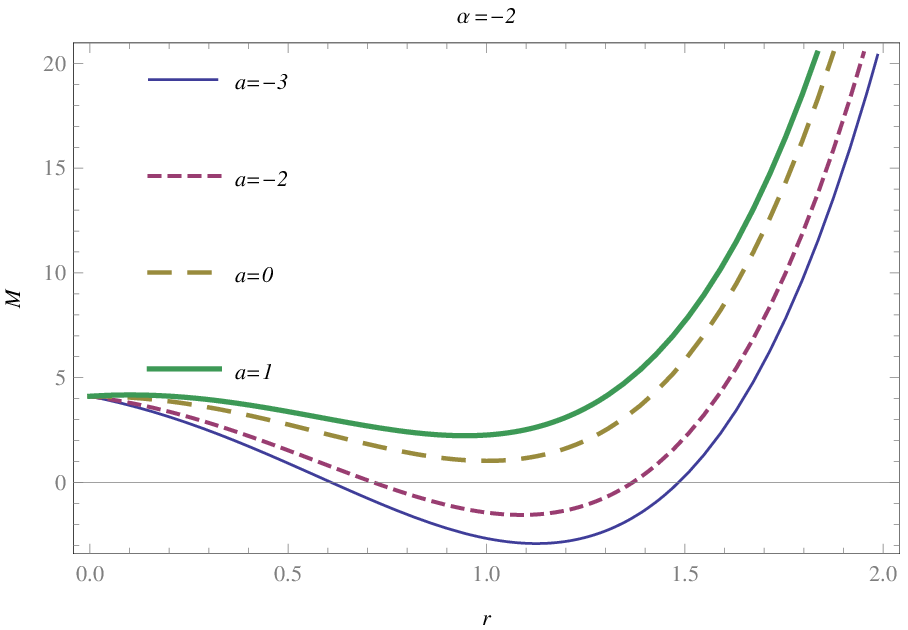} &&
    \includegraphics[scale=1.0]{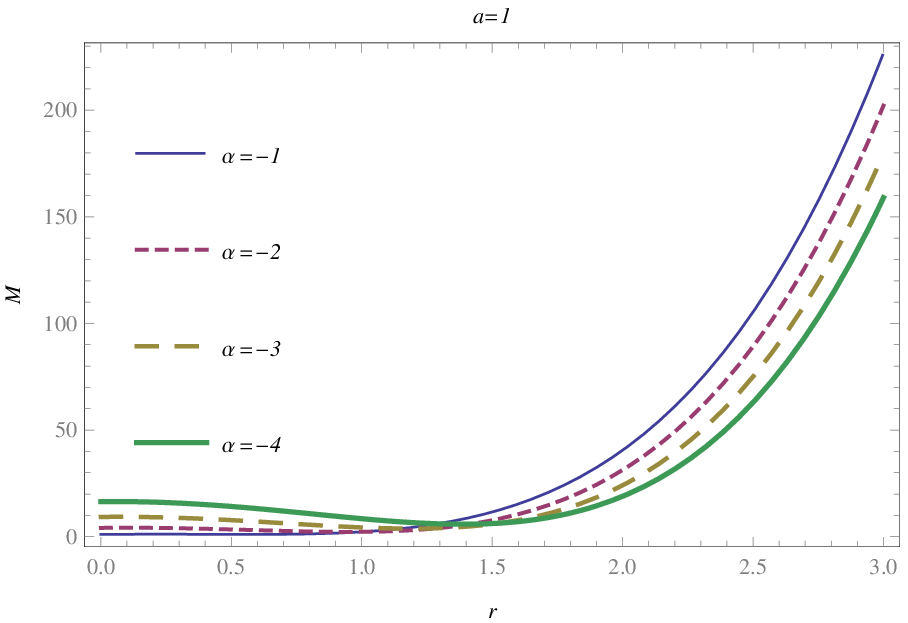}\\
    (c)&&(d)
\end{tabular}
    \end{center}
    \caption{Plots show mass $M$ as a function of a horizon radius for different values of parameters. The dotted vertical lines in (a) and (b) correspond to $r=r_0$.} \label{fig:m}
\end{figure}

In this section we present thermodynamic properties of the Lovelock black hole solution Eq.~(\ref{f-}).  As we demonstrate in the following, like any other black holes it also has thermodynamic properties. The Arnowitt-Deser-Misner(ADM) mass is defined
\begin{equation}
M=\frac{(D-2)V_{D-2}}{16\pi} \omega,
\end{equation}
where $V_{D-2}=2\pi^{(D-1)/2}\Gamma[(D-1)/2]$ is the area of a unit $(D-2)$-sphere. Thus the gravitational mass of the black hole is determined by $f(r_h)=0$, which in terms of a horizon $r_h$, from Eq.~(\ref{f-}), reads
\begin{equation}
M=  \frac{1}{16} \pi ^2 \left(2 a r_h+\frac{5}{3}\alpha ^2+5r_h^4+5 \alpha  r_h^2\right).\label{mass}
\end{equation}
As $a\to0$, Eq.~(\ref{mass}) becomes
\begin{equation}
M\to\frac{5}{16} \pi ^2 \left(\frac{1}{3}\alpha ^2+r_h^4+ \alpha  r_h^2\right),\nonumber
\end{equation}
which is found in \cite{Li:2011uk} in the limits of the vanishing
Born-Infield electromagnetic field and cosmological constant,
$(\beta,\Lambda)\to0$. Furthermore, in the limits $(a,\alpha)\to0$
it leads to $M\to\frac{5}{16}\pi^2r^4_h$, which is mass for the
Schwarzschild black hole in seven dimension \cite{Myers:1988ze}. Imposing positive mass condition gives either possible range of horizon radii or mass, for fixed $a$ and $\alpha$. If the mass function $M(r_h)$ has zeros, it will be critical points telling us which range of horizons is allowed. If the minimum of the mass function is positive it gives the minimum mass. Let us first think about the case $\alpha>0$, where a singularity is not covered by a horizon. For $\alpha>0$ and $a\geq0$ the minimum mass is $\frac{5}{16} \pi ^2\frac{1}{3}\alpha^2$. For $\alpha>0$ and $a<0$ the mass function can have two zeros, which means that it has zero minimum mass and there is no horizon between such two zeros. Next consider $\alpha<0$. If one is concerned with case that a singularity is covered by a horizon, in this case the minimum mass is $M(r_0)$, which can be found in Eq.~(\ref{omegam}).

\begin{equation}
M_m=\frac{1}{16} \pi ^2 \left[2 a (-\alpha)^{1/2}+\frac{5}{3}\alpha ^2\right].\label{Mm}
\end{equation}
In Fig.~\ref{fig:m} mass function $M(r_h)$ is plotted as a function of $r_h$ for different values of $(a,\alpha)$.

\begin{figure}[h!]
  \begin{center}
\begin{tabular}{ccc}
    \includegraphics[scale=1.0]{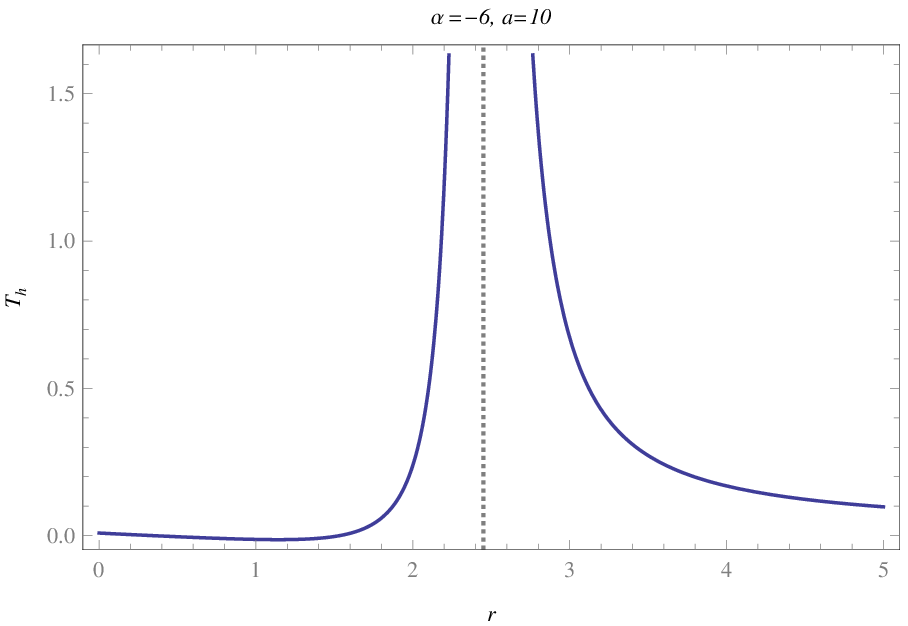} &&
    \includegraphics[scale=1.0]{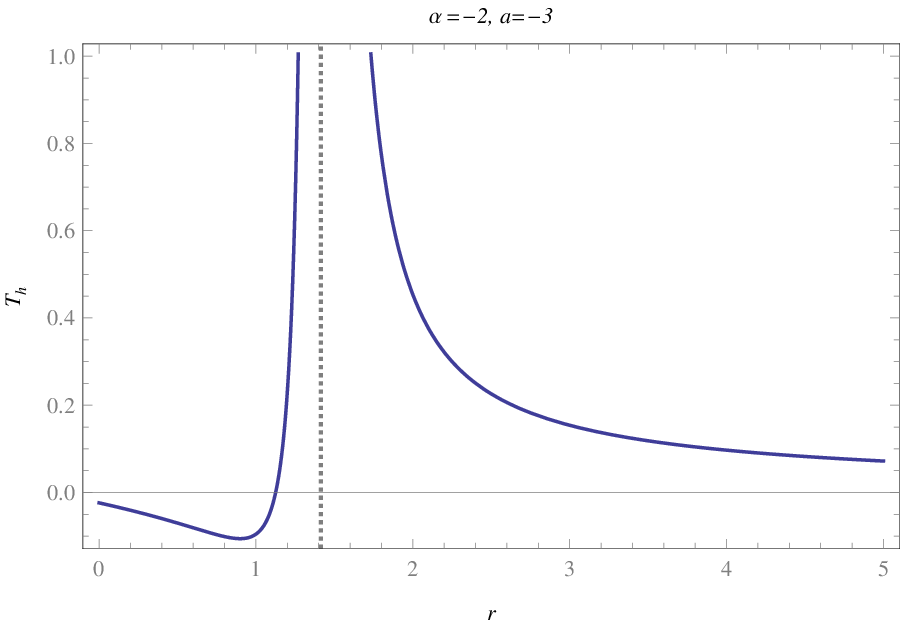}\\
    (a)&&(b)\\
    \includegraphics[scale=1.0]{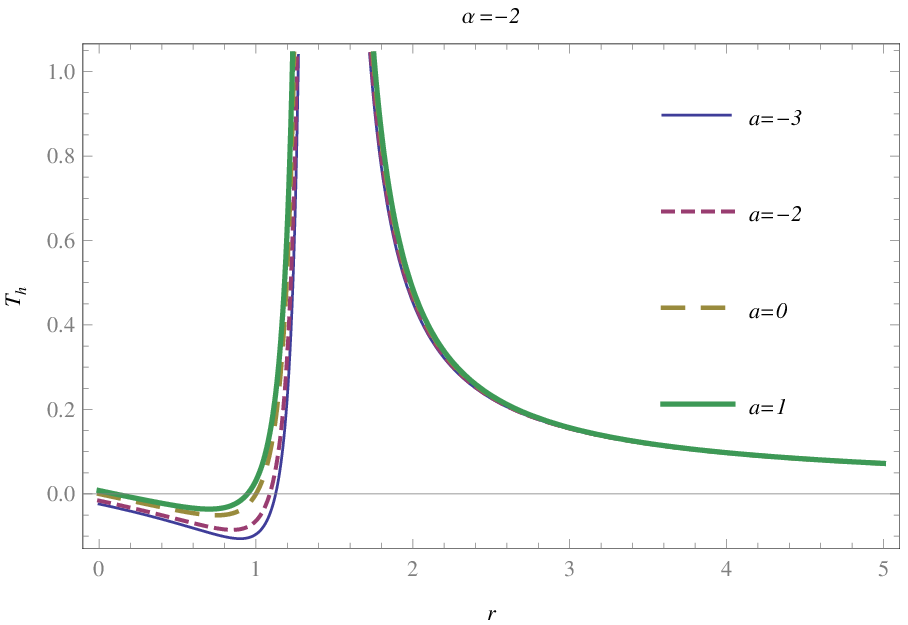} &&
     \includegraphics[scale=1.0]{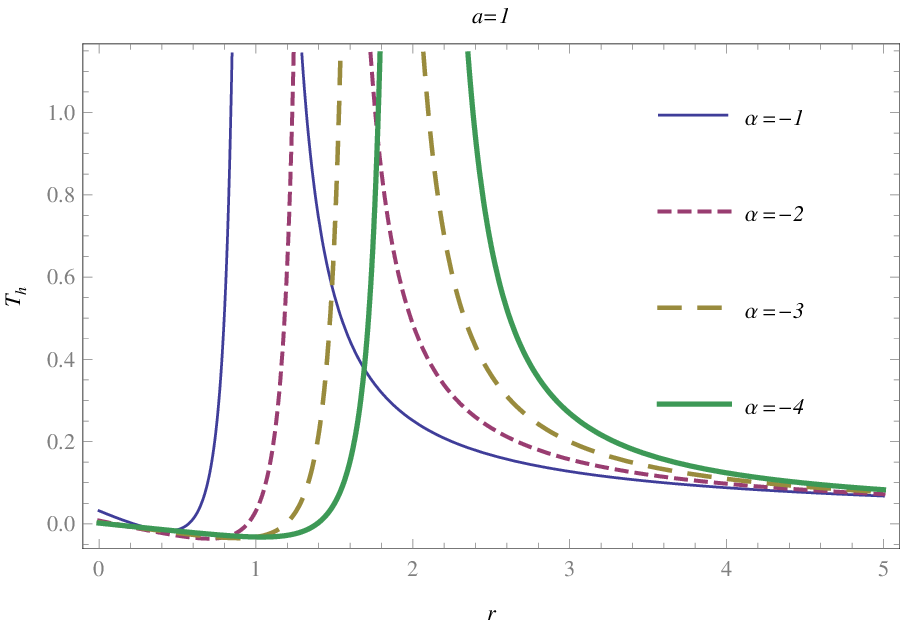}\\
    (c)&&(d)
\end{tabular}
    \end{center}
    \caption{Plots show the Hawking temperature of the black holes, $T_h$ as a function of a horizon radius for different values of parameters.  The dotted vertical lines in (a) and (b) correspond to $r=r_0$.}\label{fig:t}
\end{figure}

The Hawking temperature associated with a black hole is calculated using $T_h=\kappa/2\pi$, where $\kappa$ is a surface gravity of a horizon.  Hence, the temperature $T_h$ at the horizon can be calculated by the definition $T_h=f^\prime(r_h)/4\pi$, which is simplified to
\begin{equation}
T_h=\frac{a+10r^3_h+5 \alpha  r_h}{10\pi(\alpha +r^2_h)^2}.\label{T}
\end{equation}
From Eq.~(\ref{mass}) the positivity of mass means $ar_h>-\frac{5}{6}\alpha^2-\frac{5}{2}r_h^4-\frac{5}{2}\alpha r_h^2$. This leads to the inequality, $r_h(a+10r^3_h+5\alpha r_h)>5\left(\frac{3}{2}r_h^4+\frac{1}{2}\alpha r_h^2-\frac{1}{6}\alpha^2\right)$. One can easily see that the right hand side in the inequality is positive for $r_h>r_0$, i.e., $a+10r^3_h+5\alpha r_h>0$. Thus, $T>0$ for $r_h>r_0$. Therefore, for $r_h>r_0$ whenever mass is positive, temperature is also positive. In the limit $a\to0$, the temperature, Eq.~(\ref{T}) goes to
\begin{equation}
T_h\to\frac{2r^3_h+\alpha  r_h}{2\pi(\alpha +r^2_h)^2}, \nonumber
\end{equation}
which is found in \cite{Li:2011uk} as $(\beta,\Lambda)\to 0$. In addition the limit $\alpha\to 0$ further reduces it to the temperature for the Schwarzschild black hole, $T_h\to 1/\pi r_h$. Fig.~\ref{fig:t} plots show the Hawking temperature of the black holes for different values of parameters.
\begin{figure}[h!]
  \begin{center}
     \includegraphics[scale=1.0]{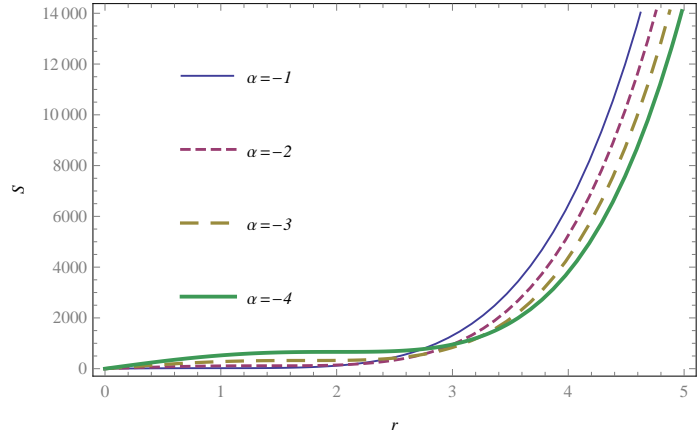}
    \end{center}
    \caption{The plot shows entropy as a function of a horizon radius with different values of $\alpha$.}\label{fig:S}
\end{figure}

For the Schwarzschild black hole in $D$ dimension, the entropy of a black hole, $S$ is given by $S=A_h/4$ with the area of the horizon, $A_h$, which is $D-2$ dimensional surface area of a sphere.  However, the black hole is supposed to obey the first law of thermodynamics $dM = TdS$. To calculate the entropy, the integral can be done with respect to $r_h$,
\begin{equation}
S=\int \frac{dM}{T}=\int T^{-1}\frac{\partial M}{\partial r_h}{dr_h}.
\end{equation}
$\frac{\partial M}{\partial r_h}=\frac{1}{16}\pi^2(2a+20r^3_h+10\alpha r_h)$ leads to
\begin{equation}
S=\int^{r_h}_{0}dr\frac{5}{4} \pi ^3 \left(\alpha +r^2\right)^2+\text{const.}
=\frac{1}{12} \pi ^3 \left[3r_h^5+10 \alpha r_h^3+15\alpha ^2 r_h\right]+\text{const.}.\label{S}
\end{equation}
Here we calculate the entropy by integration from $0$ to $r_h$. This is a usual way to define entropy in order to make entropy vanish when a horizon length becomes zero. Although a horizon length must be greater than $r_0$ it should not be a concern here because the difference is only an additive constant.  The integrand in $S$ is positive, so $S\geq0$ in any case. Also, it is worthwhile to notice that the entropy is independently expressed of $a$. A similar case happens in \cite{Li:2011uk}. Actually it can be seen that this property holds in the general Lovelock theory for any static, spherically symmetric energy momentum tensor, \cite{THLee:2015}. The entropy expression in \cite{Li:2011uk} coincides with Eq.~(\ref{S}). Using the areas of spheres $A_{n}=2\pi^{(n+1)/2}r^{n}/\Gamma[(n+1)/2]$ for an $n$ dimensional surface $A_{5}=\pi^3r^5$. Only the second term $\frac{1}{4} \pi ^3 r_h^5$ in Eq.~(\ref{S}) reflects the area law
$S=A_h/4$ and the rest are usually considered as quantum corrections in higher dimension. Fig.~\ref{fig:S} plots behaviours of the entropy in terms of a horizon radius for different values of $\alpha$.
\begin{figure}[h!]
  \begin{center}
\begin{tabular}{ccc}
     \includegraphics[scale=1.0]{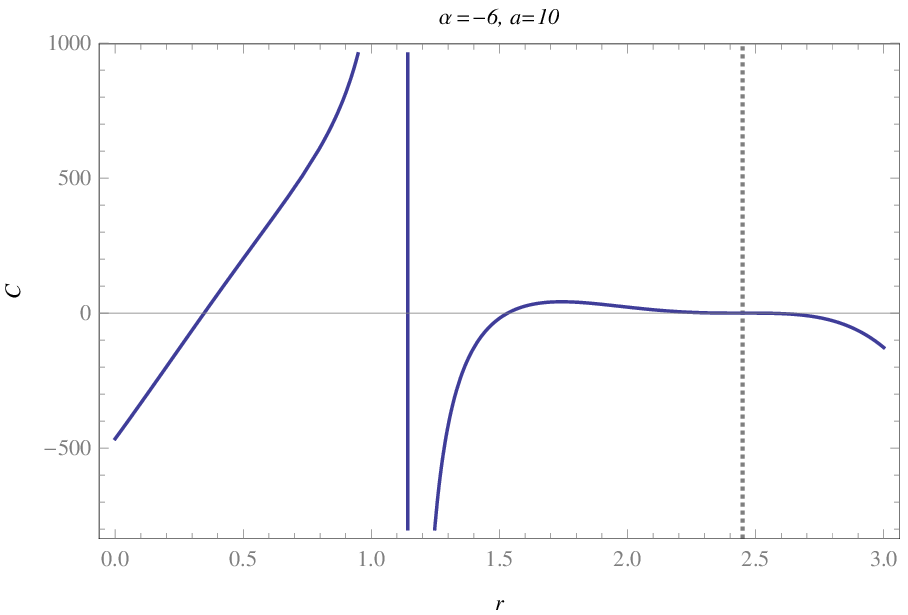} &&
    \includegraphics[scale=1.0]{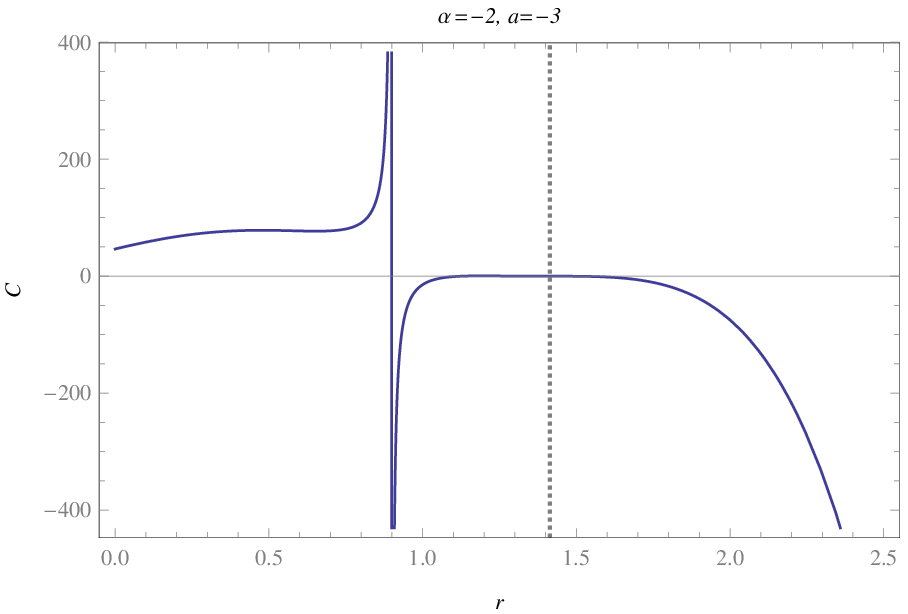}\\
    (a)&&(b)\\
    \includegraphics[scale=1.0]{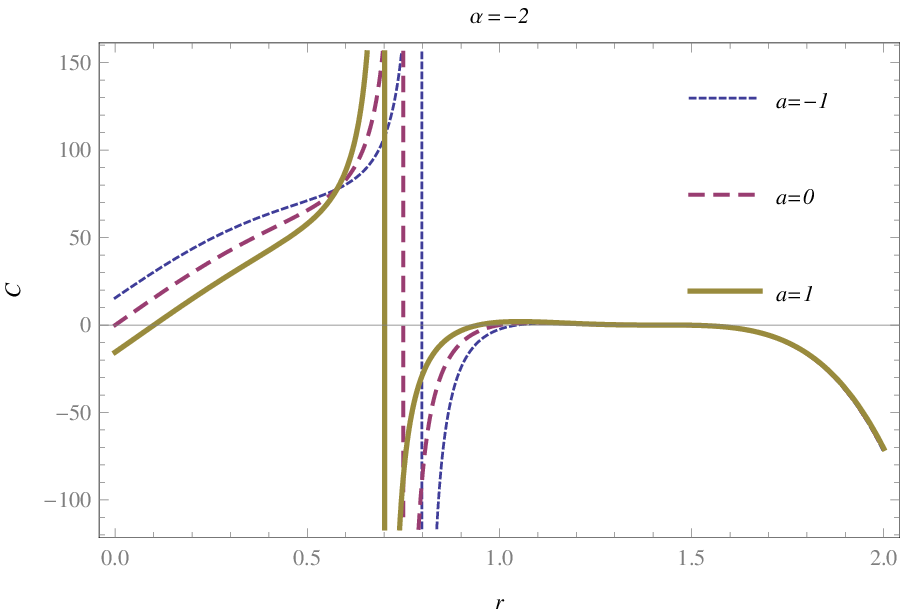} &&
     \includegraphics[scale=1.0]{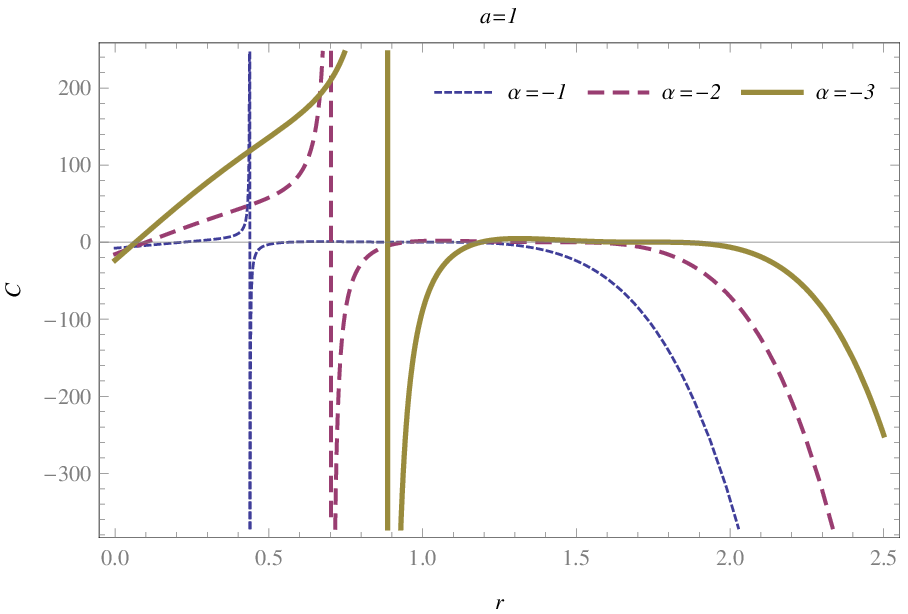}\\
    (c)&&(d)
\end{tabular}
    \end{center}
    \caption{Plots show heat capacity as a function of a horizon radius with different values of parameters. The dotted vertical lines in (a) and (b) correspond to $r=r_0$.}\label{fig:C1}
\end{figure}

The heat capacity is expressed by,
\begin{equation}
 C=\frac{\partial M}{\partial T}=\frac{\partial M}{\partial r_h}\Big/\frac{\partial T}{\partial T_h}.
 \end{equation}
  Using $\frac{\partial M}{\partial r_h}$ and $\frac{\partial T_h}{\partial r_h}=\frac{5\alpha ^2-4 a r_h-10 r^4_h+15 \alpha  r^2_h}{10\pi  \left(\alpha +r^2_h\right)^3}$, we get
\begin{equation}
C=-\frac{5 \pi ^3 \left(\alpha +r^2_h\right)^3 \left[a+5 \left(2r^3_h+\alpha  r_h\right)\right]}{4 \left[4 a r_h+5 \left(-\alpha ^2+2r^4_h-3 \alpha  r^2_h\right)\right]}.\label{C}
\end{equation}
In the limit $a\to0$, the heat capacity Eq.~(\ref{C}) goes to
\begin{equation}
C\to -\frac{5}{4}\pi^3 \frac{(\alpha+r^2_h)^3(2r^3_h+\alpha r)}{(-\alpha^2+2r^4_h-3\alpha r^2)}, \nonumber
\end{equation}
which is found in \cite{Li:2011uk} as $(\beta,\Lambda)\to0$.
We have just seen above that from the positive mass condition, $a+5(2r^3_h+\alpha r_h)>0$ for $r_h>r_0$. Using the same condition, we notice that in the denominator in Eq.~(\ref{C}), for $r_h>r_0$
\begin{equation}
4 a r_h+5(-\alpha ^2+2r^4_h-3 \alpha  r^2_h)>-\frac{25}{3}\alpha\left(r^2_h+\frac{\alpha}{3}\right).\nonumber
\end{equation}
The right hand side is always positive when $\alpha<0$ and $r_h>r_0$, which makes $C<0$, while when $\alpha>0$, $C$ can be either positive or negative.
Therefore, the heat capacity is always negative for $r_h>r_0$ and $\alpha<0$ with positive mass and hence the black hole is thermodynamically unstable in positive mass region. However, it does not necessarily mean that the black hole is unstable as the Schwarzschild black hole is stable. The heat capacity of the black holes is plotted in Fig.~\ref{fig:C1} for various values of the parameters $(a,\alpha)$. It turns out that the parameters $(a,\alpha)$ influence the thermodynamic stability of black holes. There exist transition points at which a sign of heat capacity changes, i.e. boundaries between thermodynamically stable and unstable regions.

Let us define a transition point $r_t$ and a critical point $r_c$ such that $C$ becomes zero and diverges, respectively. $r_c$ is within the positive mass region but $r_t$ does not always belong to it. $r_t$ can occurs at either $A(r_t)=(\alpha+r^2_t)^3=0$ or $B(r_t)=a+5 \left(2r^3_t+\alpha  r_t\right)=0$, which makes temperature zero. We saw that the positive mass condition guarantees $B$ is positive for $r_h>r_0$. There always exists one and only one $r_c$ for any $a$ at $D(r_c)=4a r_c+5 \left(-\alpha ^2+2r^4_c-3 \alpha  r^2_c\right)=0$.

For $\alpha>0$, we can see that $A>0$ and $r_c>r_t$ if $r_t$ exists. Signs of $C$ are changing like $(-, 0(r_t),+,0(r_c),-)$, from $-\infty$ to $\infty$.

\begin{figure}[h!]
  \begin{center}
\begin{tabular}{ccc}
     \includegraphics[scale=1.0]{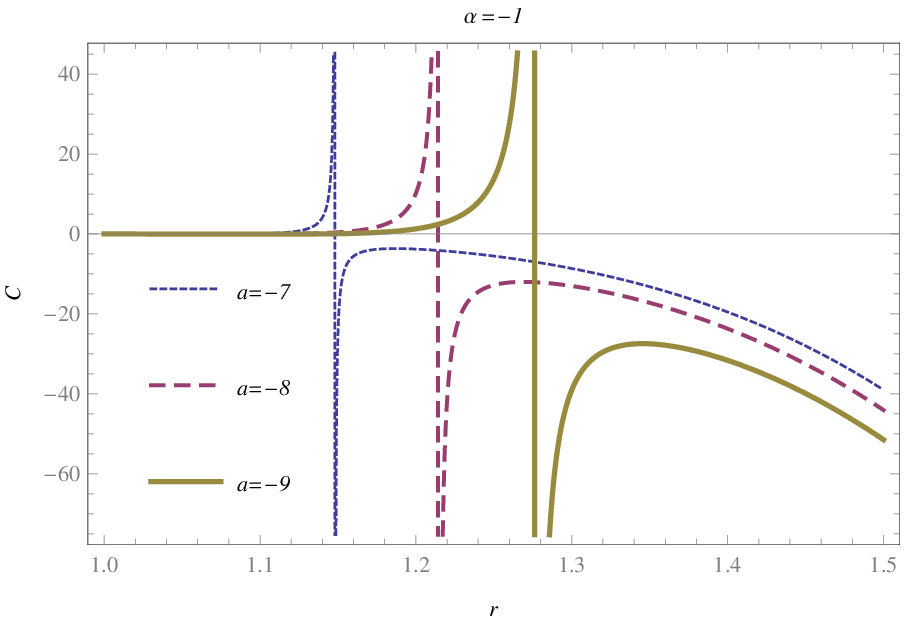} &&
    \includegraphics[scale=1.0]{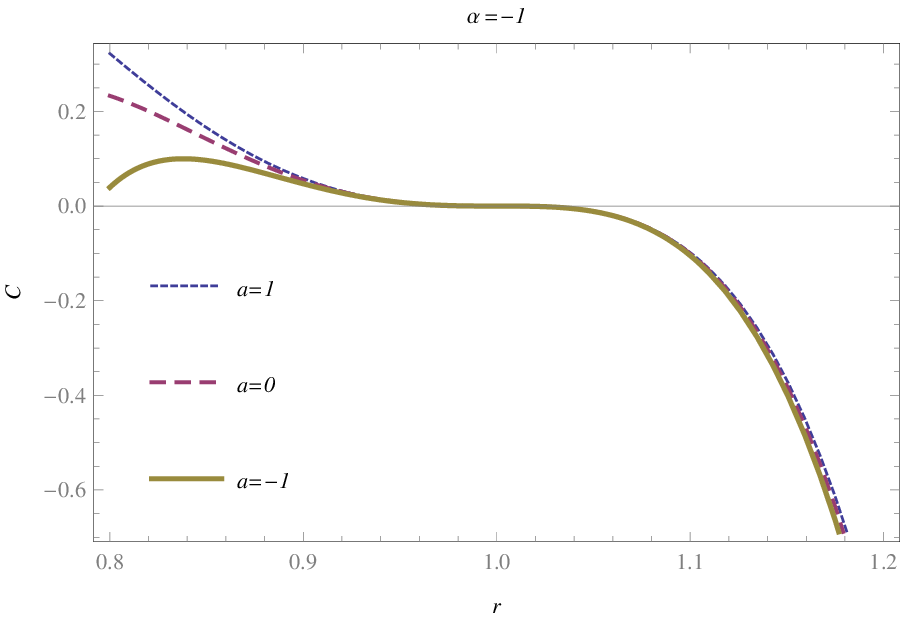}\\
    (a)&&(b)
\end{tabular}
    \end{center}
    \caption{Plots show how heat capacity changes in different ranges of $a$.}\label{fig:C2}
\end{figure}

For $\alpha<0$ when the positive mass condition imposed there is no zero in $B$ for $r_t>r_0$, which is the non-naked singularity condition, so $r_0$ is the only zero point satisfying both the positive mass and non-naked singularity conditions. Fig.~\ref{fig:C2} shows how heat capacity behaviour changes with $a$ after some critical values $a_c$. For $|a|>a_c$, $r_c>r_t$  while for $|a|<a_c$, $r_c<r_t$. For $|a|<a_c$ and $r_c\simeq r_0$ $C$ changes much less as $a$ changes for $r_h>r_0$.
\section{Conclusion}\label{Conclusion}
The Lovelock theory is a natural extension of the Einstein theory of general relativity to higher dimensions and it is of a great arena for theoretical physics research. The Lovelock theory describes string-inspired corrections of the Einstein-Hilbert action and hence admits the general relativity as a particular case. In this paper, we have obtained exact static, spherically symmetric black hole solutions to the third order Lovelock gravity in a string cloud background in seven dimensions with help of carefully choosing coefficients of the curvature correction terms, thereby generalizing the static, spherically symmetric black hole solutions for these theories. These solutions  possess rich properties of black holes and in the limits go over to black holes in Einstein's gravity.

The string cloud parameter $a$ changes the number of horizons and singularities. For $\alpha>0$, $a$ can change the maximum number of horizons and the positivity of $a$ creates a singularity. In this case a singularity, if any, is naked. For $\alpha<0$ there is an interesting property. A horizon cannot exists between $r_0=|\alpha|^{1/2}$ and a singular point $r_*$ and hence we need know only whether a horizon is ahead or behind from $r_0$ instead of $r_*$ in order to see whether a singularity is naked. There is only one singularity. For $a\geq0$ when a singularity is non-naked there should be only one horizon. For $a<0$, when a singularity is non-naked, the singularity must be between two horizons.

We proceeded to find exact expressions  for the thermodynamic quantities like the black hole mass, the Hawking temperature, entropy and heat capacity and in turn also analysed the thermodynamic stability of black holes. In addition we explicitly brought out the effect of a string cloud background on black hole solutions and their thermodynamics.

We found that a positive mass condition leads to positive temperature for $r_h>r_0$, which is a condition for a non-naked singularity for $\alpha<0$. The entropy does not depend on a string cloud. This result can be extended for any spherical and static source in the Lovelock theory \cite{THLee:2015}. We also see the entropy does not obey the horizon area formula. For heat capacity, $\alpha>0$ a critical(singular) point $r_c$ is greater than a transition (zero) point $r_t$ and for $\alpha<0$ when a positive mass and a non-naked singularity conditions applied, $r_t=r_0$ and there is no $r_c$. In this case the heat capacity is negative, telling us that the black hole is thermodynamically unstable like the Schwarzschild black hole.

The possibility of a further generalization of these results in arbitrary dimensional Lovelock gravity is an interesting problem for future research.

\begin{acknowledgements}
SGG thanks IUCAA for hospitality while part of the work was being done, and to  SERB-DST, Government of India for
 Research Project Grant NO SB/S2/HEP-008/2014.
\end{acknowledgements}

\end{document}